\documentstyle[11pt,newpasp,twoside,epsf]{article}
\def\edcomment#1{\iffalse\marginpar{\raggedright\sl#1\/}\else\relax\fi}
\reversemarginpar

\begin{document}
\title{Optical Seeing and Infrared Atmospheric Transparency in the Upper
Atacama Desert}
 \author{Riccardo Giovanelli}
\affil{Cornell University, Dept. of Astronomy, Ithaca, NY 14853, USA}

\begin{abstract}
With plans to build a large optical/IR telescope in the Atacama Desert,
a site survey campaign has been under way since 1998 to characterize the 
optical seeing and the IR transparency in the Chajnantor Plateau region
in Chile. Results of that campaign are herein presented.
\end{abstract}

\section{Introduction}

The Atacama Desert in northern Chile is one of the driest regions on Earth. It 
lies between the Coastal Cordillera to the west and the Andes to the east. The 
region of the Altiplano to the east of the Salar de Atacama known as Llano de 
Chajnantor, a plateau of altitude near 5000 m, was selected by the U.S. National 
Radio Astronomy Observatory (NRAO) as the future site for its Millimeter Array 
Project, while a neighboring plateau, {\it Pampa La Bola}, was selected by the 
Nobeyama Radio Observatory (NRO) of Japan for its Large Millimeter and Submillimeter 
Array (LMSA) Project. These sites are within a few km from each other (see Fig. 
1), at elevations between 4800 and 5050 m above mean sea level. 
Successively, the Millimeter Array Project has evolved into a U.S.--Europe 
consortium to build the Atacama Large Millimeter Array (ALMA), which may also
be joined by Japan. Other radio
and optical Astronomy consortia are under development for operation in this 
region, which has the potential for expanding into a major world astronomical
center. Cerro Paranal, the site of the European Southern Observatory's (ESO) Very
Large Telescope Project, is about 300 km to the southwest, on a peak on the
Pacific Coastal Cordillera.

The climatic qualities that make the Atacama region especially attractive
to astronomers extend over a latitudinal band a few hundred kilometers wide, 
about the Tropic of Capricorn. The access to good quality services and good 
communications further focuses attention on the regions in the vicinity of the 
cities of Antofagasta, Calama and the village of San Pedro de 
Atacama. The presence of the VLT and the likely establishment of major national 
and international research centers, such as ALMA, adds promise of 
scientific and operational synergism to the region. The government of Chile has
legislated the protection of an area which includes the Llano de Chajnantor, 
the Pampa La Bola and the surrounding peaks, as a {\it National Science Preserve}.

\begin{figure}  
\plotone{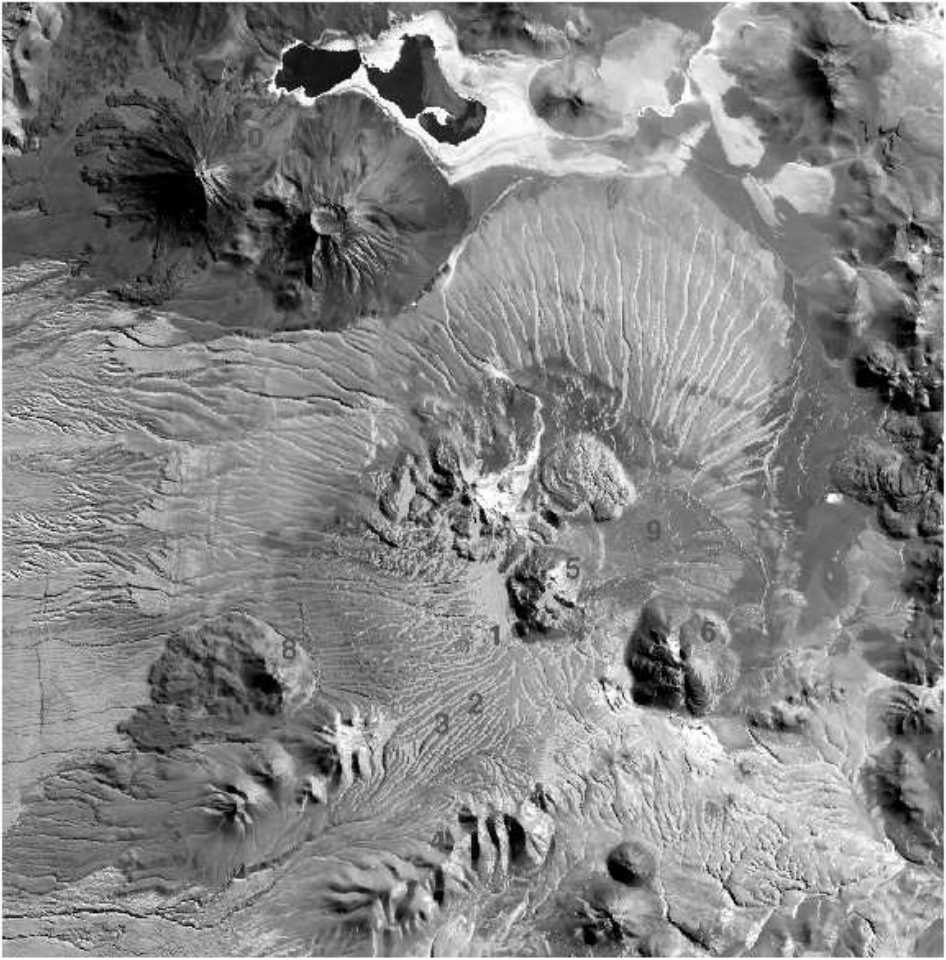}
\caption{Satellite view of the Chajnantor Plateau region. 
The image subtends  approximately 40 km to the side.
North is up and left is West. The labeled features, as referred
in the text are (elevations in parenthesis): 1. Cerro Chico (5150 m);
2. ALMA test site (5050 m); 3. Sonde launch (5030 m); 4. Cerro Honar 
(5400 m); 5. Cerro Chajnantor (5650 m); 6. Cerro Chasc\' on (5750 m);
7. Cerro Toco (5650 m); 8. Cerro Negro (5150 m); 9. Pampa La Bola,
NRO test site (4800 m); 10. Cerro Licancabur, Bolivian border (5950 m).
\label{satview}}
\end{figure}

Cornell University, jointly with U. of Texas and U. of Virginia, has initiated
a project to build a 15--m class optical/IR telescope in the region. Since 
mid--1998, a survey campaign has been under way to characterize the optical 
and IR astronomical properties of the sites in the Chajnantor Plateau region.
Here we report on preliminary results, particularly regarding seeing and water 
vapor measurements.

\section{Weather at Chajnantor}

Weather conditions have been monitored for several years at the NRAO, ESO and NRO 
testing stations in the Chajnantor region. In spite of the altitude, those conditions 
are not extreme. The median temperature at 5000 m is $-2.6$ C; the amplitude of the 
diurnal cycle is about 13 C and that of the annual cycle is about 11 C. Wind speeds 
are high during the day, but at the plateau level they drop drastically soon after 
sunset. Details can be seen in the NRAO and ESO websites. At peaks in the plateau 
region, however, circumstances are less benign, as shown in Figure 2, which 
displays median profiles of wind speed and temperature, as a function of elevation 
above the plateau, obtained from radiosonde data as mentioned below. Temperature 
changes little in a few hundred m height, but while at night the wind speed at 
5000 m drops typically below 
4 m/s, at elevations of only a few hundred m above the plateau, median speeds near 
10 m/s are encountered. These speeds are safely within the operational limits at 
modern observatories; however testing {\it in situ} of gusting conditions will be 
necessary.

\begin{figure}    
\plottwo{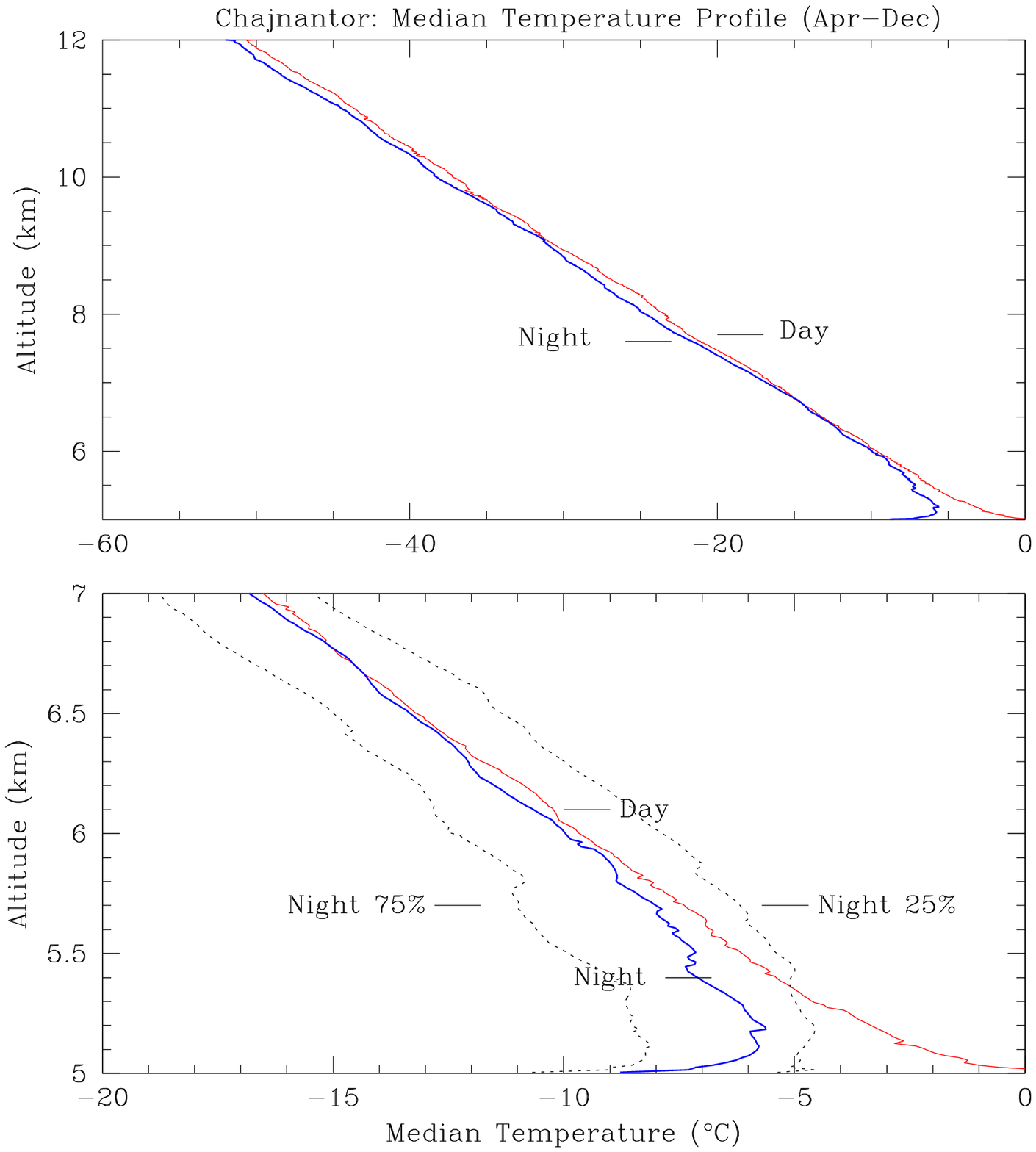}{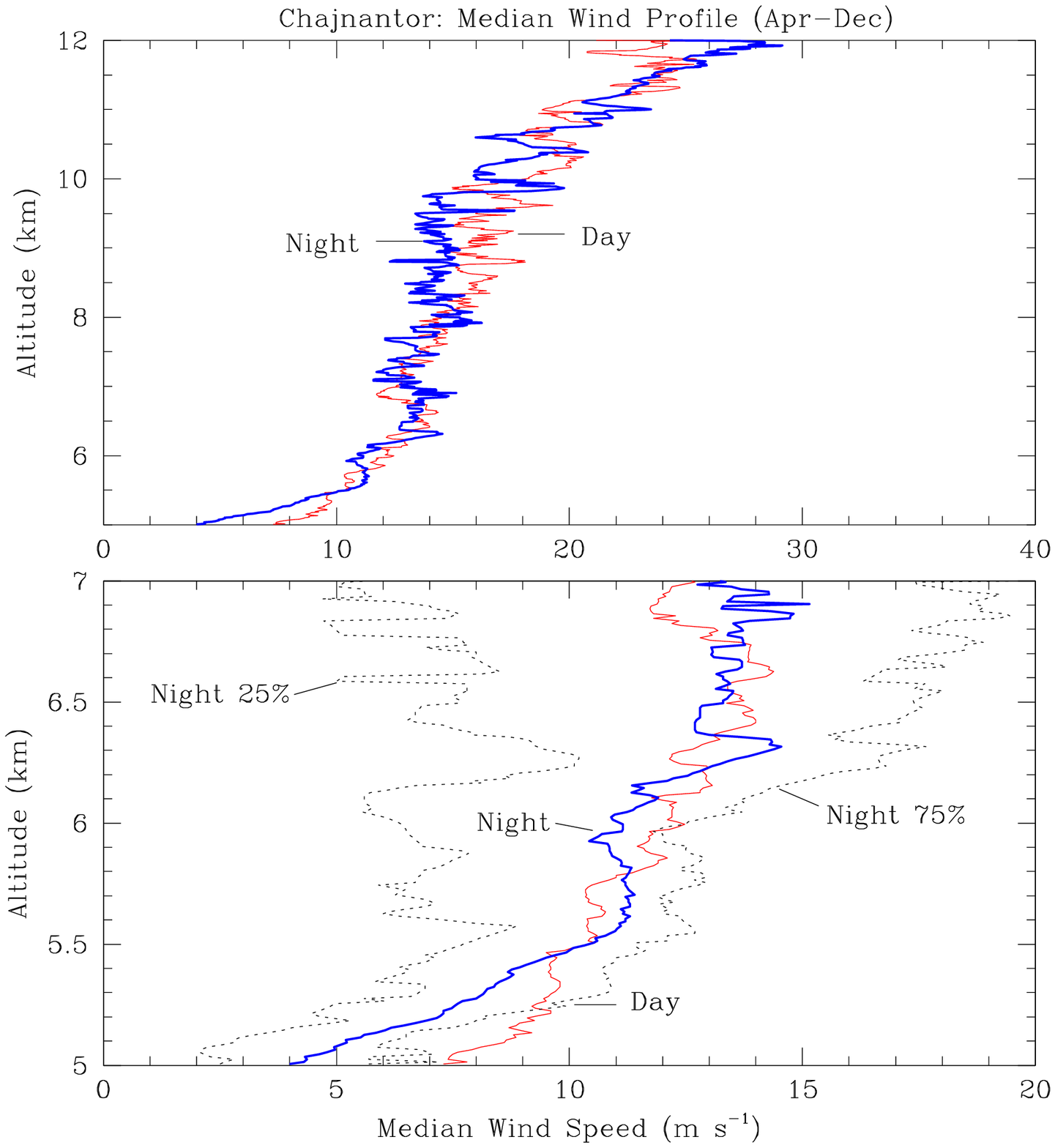}
\caption{Median temperature (left) and wind speed profiles above Chajnantor.
Night profiles derived from 30 sonde flights; day profiles obtained from 65
sonde flights. Dotted lines indicate the 25\% and 75\% quartiles during the
night. Lower panels zoom on the lower atmosphere. Sondes weare launched 
between April and December (of the 30 night flights,
14 were in November).
\label{meds}}
\end{figure}

As discussed by Erasmus (2000) in these proceedings, the Chajnantor region
enjoys a high fraction of clear nights. he reports a fraction higher than 90\% 
of nights with limited or no cloud cover above 7500 m elevation in 1995. Our 
visual records of 84 nights between May 1998 and October 2000 yields a fraction 
of 63\% photometric nights, and an additional 18\% deemed astronomically useful 
but not photometric.

\section{Optical Seeing}

We aim to establish the statistical properties of seeing at the best sites in
the National Science Preserve region and vicinity, in Atacama. A seeing campaign
startegy was develop that would follow three phases:

\noindent I. We would first establish a reference frame for seeing statistics 
in the region at an easily accessible site, possibly above the local boundary 
layer. We would determine reliably the site's average seeing properties and 
seasonal variations, by carrying out a series of seeing runs spread over the 
seasonal cycle. This phase requires the deployment of a single seeing monitor.

\noindent II. Next, we would compare 
characteristics of potentially attractive sites by means of relatively 
brief runs of seeing measurements at those sites, simultaneous with 
measurements at the reference site. This phase requires the deployment of
two seeing monitors.

\noindent III. Finally, once the site with the best comparative characteristics 
is identified, a long--term campaign of continuous monitoring with a robotic 
seeing monitor would be carried out.

Given its accessibility, partial emergence above the atmospheric 
boundary layer and central location in the Science Preserve Area, we chose 
Cerro Chico as the site at which a seeing reference standard for the region 
would be established. So far, we have carried out seven runs of 8--10 days duration
each, spaced by approximately 2--3 months.

We used a Differential Image Motion Monitor (DIMM) obtained on loan from ESO.
It operates at a wavelength of 0.5 $\mu$ and as described by Sarazin \& Roddier
(1990). It consists of an 11" Celestron telescope with two 8 cm apertures 
separated by 19.2 cm. Data taking with the device alternates images of 10 ms and 
20 ms exposure; pairs of 10 ms and 20 ms images are used to extrapolate to an
ideal ``zero exposure seeing'' figure, which would ``freeze'' the flow of atmospheric
turbulence past the telescope apertures. During operation, the DIMM apertures
stand some 2.5 m above the ground. As a result, some fraction of ground boundary 
layer turbulence contributes to the seeing values thus obtained. This contribution
may not be larger than 0.2" (Martin et al. 2000). For simultaneous operation at
two different sites, we have built a second device, identical to the ESO DIMM,
which we started deploying in April 2000. A robotic DIMM is under construction.

Our first seeing run, of May 1998, was carried out near the NRAO testing site
(``ALMA Container'', position labeled `2' in Fig. 1), at an elevation
of 5050 m. This decision was made in order to ascertain our and the equipment's
ability to effectively function at high altitude; caution advised that such
verification be made near a shelter. Successively, all runs discussed here were
carried out a the summit of Cerro Chico, some 2 km NW and about 100 m higher
than the ALMA Container.

\begin{figure}   
\plotone{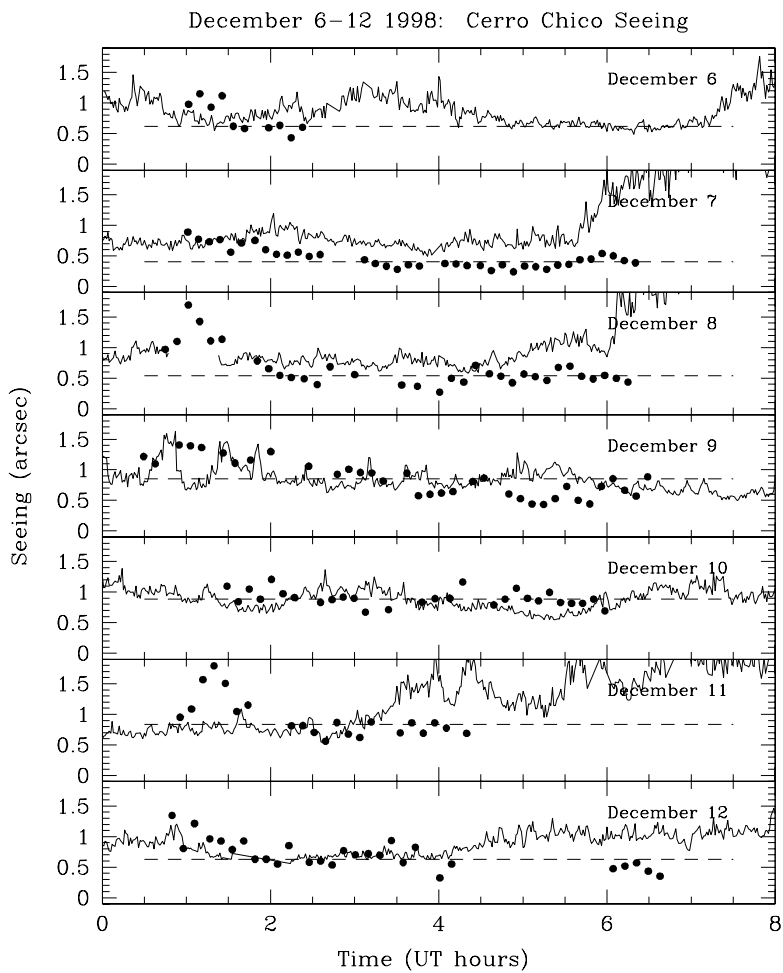}
\caption{Time series of ``zero exposure FWHM seeing'' measurements for the 
December 1998 run. Filled circles are 8--min seeing averages at Cerro Chico;
thin solid lines represent simultaneous seeing at Cerro Paranal. Dashed lines
are nightly Cerro Chico medians. Analogous plots for other runs can be seen
at {\it http://www.astro.cornell.edu/atacama/}. Midnight is at 4:31 hrs UT.
\label{see}}
\end{figure}

Figure 3 displays the median ``zero exposure seeing'' data for each day 
of the December 1998 run (other runs were in July, October 1998, March 1999,
April, July and October 2000). The median FWHM seeing for the December 1998 run was
0.66". {\it The median for the 38 nights of Cerro Chico measurements, July 1998 to
October 2000, was 0.71". By comparison, the median for the May 1998 run at the
ALMA Container was 1.09". The median seeing at Cerro Paranal, for the same nights
for which we have data from Cerro Chico, was 0.80" (measured with a similarly
calibrated DIMM). The 10ms and 20 ms median seeing (i.e. values not extrapolated
to zero exposure) for Cerro Chico were respectively 0.61" and 0.52"}.

Note that the seeing at Cerro Paranal during the Cerro Chico survey was markedly
worse than the Paranal historical median (0.66", or 21\% lower than the value of 
0.80" of the ``Chico nights''). This was noted by Sarazin \&
Navarrete (1999), who ascribed it to an exceptional {\it El Ni\~ no/La Ni\~ na}
anomaly, which would have affected the Cerro Chico seeing as much as it it did
Paranal's.

In summary, we have obtained a reliable reference frame for measurements of
seeing in the Chajnantor region. Provided by observations at Cerro Chico over
38 nights spread over the year, such reference frame is currently being used
to compare different neighboring sites. While Cerro Chico does not qualify as
a site expected to deliver the best observing conditions in comparison to
other astronomically attractive sites in the region, it has easy access, it
produces results of repeatable statistics and is thus a good choice for
benchmarking quality comparisons through simultaneous measurements at different
sites. Perhaps surprisingly --- for the summit of Cerro Chico at 5150 m may still
be significantly affected by boundary layer turbulence and katabatic winds from
Cerro Chajnantor --- it also turns out to deliver high quality seeing, 
comparable with that of the best observatories on Earth. This fact allows for
good expectations for the quality of other sites in the region. The next phase
of our campaign involves a series of simultaneous seeing runs
at Chico and other summits, in the course of which we will learn the differential
qualities of the various sites, and wil be able to tie those to the increasingly
robust data sample for Cerro Chico itself.

\section{Precipitable Water Vapor and Infrared Transparency}

Since October 1998, a radiosonde launch facility has been operated at the 
Chajnantor Plateau jointly by Cornell, NRAO, ESO, SAO and NRO. In
these proceedings, Butler (2000) reports on results from the sonde launch
campaign. We have used data from 108 sondes to obtain vertical profiles of
atmospheric variables, which can help us estimate the conditions at high
elevation sites which, in the absence of roads, cannot currently be easily
reached. Most interestingly, we verify the occurrence of patterns in the
vertical distribution of precipitable water vapor (PWV), which make sites
even a few hundred m higher than the plateau quite attractive because of 
their infrared transparency.

The PWV at the plateau level, as indicated by radiosonde, tipping radiometry
at 225 and 183 GHz and FT spectroscopy, varies around a median level of about
1.2 mm. There are clear diurnal and annual variations: driest conditions are
found between late March and early December, and between local midnight and 11 
a.m. Median PWV during winter nights is significantly lower than 1 mm at the
Chajnantor Plateau. 

\begin{figure}   
\plotone{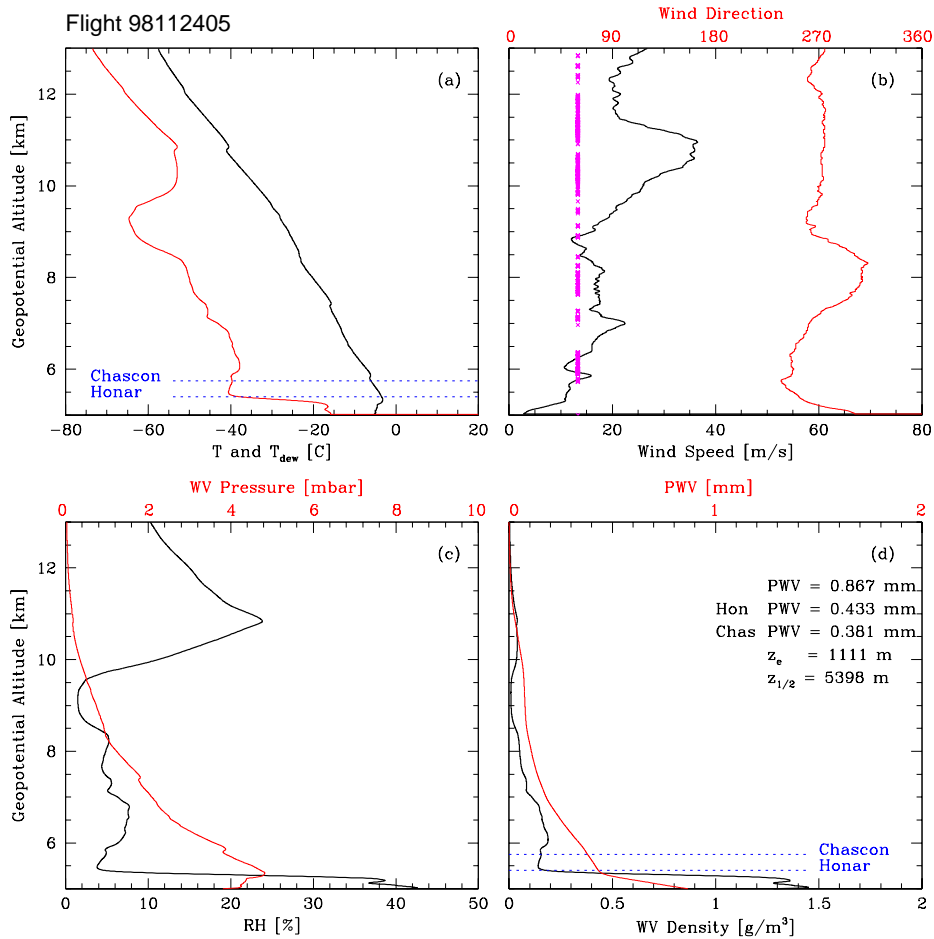}
\caption{Radiosonde data showing a case of temperature inversion below 5400 m.
Panels display: at upper left, temperature (thick) and dewpoint temperature;
at upper right, wind speed (thick) and direction; at lower left, relative humidity
(thick) and water vapor pressure; at lower right, water vapor density (thick) and
PWV. The horizontal dotted lines are at the elevations of Cerro Honar (5400 m) and 
Cerro Chasc\' on (5750 m). Note the T inversion at 5.3 km (the ground is at 5.0 km),
above which the water vapor density drops steeply. This launch took place on
November 24, 1998, at 5$^h$ UT.
\label{sonde}}
\end{figure}

The water vapor density profile above the plateau is quite variable. While the
median scale height exceeds 1000 m, temperature inversions form often below the 
elevation of local summits (say, below 5600 m), and a large fraction of the PWV 
is trapped below them. This is illustrated in the example shown in Figure 4.
Often, PWV above some of the peaks surrounding the Chajnantor Plateau, which
may be above the temperature inversion, can be very low. We estimate the winter night
median PWV in the free atmosphere above 5400 m (e.g. the elevation of Cerro Honar)
to be $\sim 0.50$ mm, with a first quartile of $\sim 0.30$ mm. Above 5700 m
(e.g. the elevation of Cero Chasc\' on), the winter night median may fall near
0.35 mm. While local topography will have an impact, the PWV at summit locations
is expected to approximate that of the free atmosphere at the same elevation.

While the results presented here are preliminary, they nonetheless suggest that
exceptional possibilities for IR and submm astronomical observations
from the ground exist in the Chajnantor Plateau region. The combination of low
water vapor and excellent seeing allow for low atmospheric background in the near 
and mid--IR. At a telescope sited on a summit in the vicinity of the Chajnantor 
plateau, numerous atmospheric windows would appear in the mid IR, up to about 
50 \micron. In the far IR and submillimeter regime, 
the 350 \micron, 450 \micron, 600 \micron, 750 \micron ~and 870
\micron ~windows reach exceptional transparency, while two useful windows 
appear near 200 \micron.

\section{Conclusions}

Preliminary results indicate that the 
Chajnantor region offers excellent astronomical observing conditions 
in the optical and IR. The region is isolated yet easy to reach,
and high quality services are available nearby. We can expect to find: 
median optical FWHM seeing approaching values of 0.5"--0.6", a 
percentage of photometric nights of 65\% or better, of astronomically 
useful nights in excess of 80\% and median PWV well below 1 mm (with 
best quartile below 0.5 mm). We will have to contend with the important
limitations --- physiological, operational, etc. --- imposed by the
high altitude environment.

\vskip 0.2in
\noindent
{\bf Acknowledgements:} The work summarized here has been carried out in
collaboration with J. Darling, M. Sarazin, J. Yu, P. Harvey, C. Henderson,
W. Hoffman, L. Keller, D. Barry, J. Cordes, S. Eikenberry, G. Gull, J. Harrington,
J.D. Smith, G. Stacey and M. Swain. Invaluable help has been provided by E.
Hardy, A. Otarola, S. Radford. It is a pleasure to acknowledge the support
of NRAO, ESO, the National Science Foundation and, most of all, Marc Sarazin.


\begin{references}
\reference Butler, B. 2000, these proceedings
\reference Erasmus, A. 2000, these proceedings
\reference Martin, F. et al. 2000, \aaps ~144, 39
\reference Sarazin, M. \& Roddier, F. 1990, \aap ~227, 294
\reference Sarazin, M. \& Navarrete,J. 1999,  {\it ESO Messenger} 97, 8
\end{references}
\end{document}